\documentclass[sigconf,nonacm]{acmart}
\usepackage{xspace}
\usepackage{multirow}
\usepackage{mathtools}
\usepackage{caption}

\usepackage{subfigure}

\usepackage{boxedminipage}
\usepackage{latexsym}
\usepackage{amsmath,algorithm}
\usepackage{bm}

\AtBeginDocument{%
  \providecommand\BibTeX{{%
    \normalfont B\kern-0.5em{\scshape i\kern-0.25em b}\kern-0.8em\TeX}}}

\setcopyright{acmcopyright}
\copyrightyear{2018}
\acmYear{2018}
\acmDOI{XXXXXXX.XXXXXXX}

\acmConference[WSDM '23]{Make sure to enter the correct
  conference title from your rights confirmation email}{Feb 27-- March 03,
  2023}{Singapore}
\acmPrice{15.00}
\acmISBN{978-1-4503-XXXX-X/18/06}

\RequirePackage[textsize=scriptsize]{todonotes}

\DeclareMathOperator{\Var}{Var}
\newtheorem{claim}{Claim}[]
\newtheorem{claimapx}{Claim}[]

\newtheorem{ass}{Assumption}[]

\newcommand{\E}{\mathbb{E}}

\begin{document}

%%
%% The "title" command has an optional parameter,
%% allowing the author to define a "short title" to be used in page headers.

%%
%% The "author" command and its associated commands are used to define
%% the authors and their affiliations.
%% Of note is the shared affiliation of the first two authors, and the
%% "authornote" and "authornotemark" commands
%% used to denote shared contribution to the research.

\def\ztitle{Privacy Aware Experiments without Cookies}
\title{\ztitle}

\author{Shiv Shankar}
\affiliation{%
    \country{\small{University of Massachusetts, USA}}
%\country{\small{USA}}
}
\email{sshankar@cics.umass.edu}

\author{Ritwik Sinha}
\affiliation{%
    \country{\small{Adobe Research, USA}}
%\country{\small{USA}}
}
\email{risinha@adobe.com}

\author{Saayan Mitra}
\affiliation{%
    \country{\small{Adobe Research, USA}}
%\country{\small{USA}}
}
\email{smitra@adobe.com}

\author{Viswanathan Swaminathan}
\affiliation{%
    \country{\small{Adobe Research, USA}}
%\country{\small{USA}}
}
\email{vishy@adobe.com}

\author{Sridhar Mahadevan}
\affiliation{%
    \country{\small{Adobe Research, USA}}
%\country{\small{USA}}
}
\email{smahadev@adobe.com}

\author{Moumita Sinha}
\affiliation{%
    \country{\small{Adobe Inc.  , USA}}
%\country{\small{USA}}
}
\email{mousinha@adobe.com}

%%
%% By default, the full list of authors will be used in the page
%% headers. Often, this list is too long, and will overlap
%% other information printed in the page headers. This command allows
%% the author to define a more concise list
%% of authors' names for this purpose.
\renewcommand{\shortauthors}{Shankar, et al.}

%%
%% The abstract is a short summary of the work to be presented in the
%% article.
\begin{abstract}
%\ritwik{Update abstract}
Consider two brands that want to jointly test alternate web experiences for their customers with an A/B test. Such collaborative tests are today enabled using \textit{third-party cookies}, where each brand has information on the identity of visitors to another website. With the imminent elimination of third-party cookies, such A/B tests will become untenable. We propose a two-stage experimental design, where the two brands only need to agree on high-level aggregate parameters of the experiment to test the alternate experiences. Our design respects the privacy of customers. We propose an estimator of the Average Treatment Effect (ATE), show that it is unbiased and theoretically compute its variance. Our demonstration describes how a marketer for a brand can design such an experiment and analyze the results. On real and simulated data, we show that the approach provides valid estimate of the ATE with low variance and is robust to the proportion of visitors overlapping across the brands. 

\end{abstract}

%%
%% The code below is generated by the tool at http://dl.acm.org/ccs.cfm.
%% Please copy and paste the code instead of the example below.
%%

%%
%% Keywords. The author(s) should pick words that accurately describe
%% the work being presented. Separate the keywords with commas.

%% A "teaser" image appears between the author and affiliation
%% information and the body of the document, and typically spans the
%% page.

%\received{20 February 2007}
%\received[revised]{12 March 2009}
%\received[accepted]{5 June 2009}

%%
%% This command processes the author and affiliation and title
%% information and builds the first part of the formatted document.
\maketitle

\section{Introduction}
%Consider two brands that want to jointly test alternate web experiences for their customers with an A/B test. Such collaborative tests are today enabled using \textit{third-party cookies}, where each brand has information on the identity of visitors to another website, ensuring a consistent treatment experience. With the imminent elimination of third-party cookies, such A/B tests will become untenable. We propose a two-stage experimental design, where the two brands only need to agree on high-level aggregate parameters of the experiment to test the alternate experiences. Our design respects the privacy of customers. We propose an unbiased estimator of the Average Treatment Effect (ATE), and provide a way to use regression adjustment to improve this estimate. On real and simulated data, we show that the approach provides valid estimate of the ATE and is robust to the proportion of visitors overlapping across the brands. Our demonstration describes how a marketer can design such an experiment and analyze the results.  

A/B testing (randomized experiment) is the gold standard for optimizing customer experiences on the web. Cookies have historically played an important role in ensuring that the same customer receives a consistent (or \emph{sticky}) experience, thus ensuring the validity of the A/B test. Consider the A/B testing scenarios, where different brands collectively test alternate customer experiences. Some examples include brands under a hotel chain, different franchises under a sporting league, related clothing brands jointly owned by a holding company, or multiple departments of a government. Thus far, such A/B tests have relied on \emph{third-party cookie}, which requires the first brand's cookie to be available on the second brand's website. This is the only mechanism to ensure that the same individual receives the same experience across the web properties of the two different brands.

\begin{figure}[t]
\centering
\includegraphics[width=1.0\linewidth]{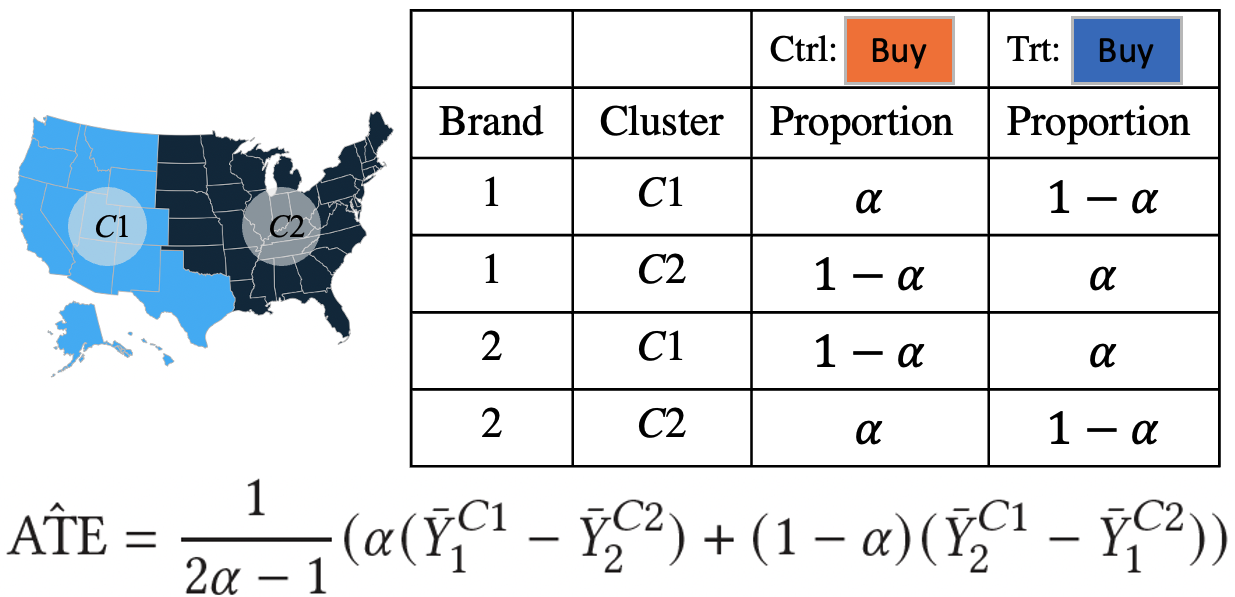}
\caption{Proposed experimental design for two brands to estimate the average treatment effect (ATE) without third party cookies. The brands only agree on a definition of clusters ($C1$ \& $C2$), the two treatments (orange and blue ``buy" buttons), and agree to randomize in proportions ($\alpha$ \& $(1-\alpha)$) swapped across brands and clusters. \label{fig:design}}
\end{figure}

A recent trend within the web ecosystem is the increased focus on customer privacy. This is embodied in laws like the General Data Protection Regulation (GDPR)\footnote{\url{https://gdpr-info.eu/}}, which require explicit permission from the visitor for tracking web sessions and allows visitors to request that their tracking data be deleted. Another manifestation of this is that all major browser ecosystems are discontinuing third-party cookies \citep{googleprivacy}. This presents significant challenges for two brands that want to jointly optimize customer experiences \citep{TheCompl75:online}, since they can no longer serve the customer a consistent web experience.

Formally, this is our question of interest. Consider a binary treatment deployed across two websites; for example, Brands 1 and 2 have jointly decided to optimize some aspect of the user experience on their websites (e.g., the colour of the buy button, a new recommendation algorithm or the discount offer \citep{sinha2020designing}). Let's call the treatments $1$ and $2$. A user may visit one or both websites. In the ideal case, a user assigned to the treatment $1$ group on the first website will also see treatment $1$ on the second website. However, in the case where the user's identity cannot be shared across the two websites (due to a lack of third-party cookies), the user is not guaranteed to see the same treatment across both websites. This leads to four potential treatment exposures, leading to the potential outcomes $Y_{11}$, $Y_{21}$, $Y_{12}$, and $Y_{22}$ (first and second positions denoting the treatments at the first and second websites). The potential outcomes that are of real interest to the two brands are $Y_{11}$ and $Y_{22}$. The question of interest is whether the treatment has any effect, which in the world with third-party cookies corresponds to the effect $E[Y_{11} - Y_{22}]$. This problem would be easily solved if the websites could share information about which users were exposed to which treatment. However, it is not clear whether one can estimate this effect without sharing such individual user-level information. 

In our work, we present a multi-stage randomization design that allows the estimation of the desired effect without sharing any user-level information. Figure \ref{fig:design} describes our proposed design. The two brands only agree on two aggregate parameters of the test: 1) the two treatments being tested, (2) a notion of clusters (dividing the population into $C1$ and $C2$, e.g., geographies, time, device types), and (3) the nuisance parameter $\alpha$ ($\neq 1/2$). No individual or identity data needs to be shared between the brands, thus respecting customer privacy. First, we show that our proposed average treatment effect (ATE) is unbiased. Next, we further propose a way of performing regression adjustment, which further helps statistical power by using attributes of the web visitors. Finally, with experiments on both simulated and real-world web data, we show that our estimate has a lower bias than the naive ATE (difference of treatment means). In our demo, we show how a marketer for a brand can use our design in a two-stage test and analyze the results.

%In summary we can state that we want to infer treatment effects from 2-component treatments when the per outcome-unit allocation of treatment components is unknown.
%\ritwik{Add contributions}

\section{Related Work}

The problems of using cookie-level identifiers on the web as a proxy for the individual's true identity is well studied \citep{chatterjee2003modeling, manchanda2006effect, rutz2011modeling, bleier2015personalized, hoban2015effects}. But looking beyond cookies as the identity in digital marketing needs to be better-studied \cite {thomas2021planning}. The Privacy Sandbox \citep{googleprivacy} is an initiative with proposals tackling privacy-related challenges for ad-targeting, delivery, and measurement. These include proposals like FLoC and TOPICS, but they are all early proposals subject to changes and uncertainty.
%Customer data platforms (CDPs) are also now of renewed interest to advertisers as they use personally identifiable information (PII). However with increasing privacy concerns, there is also concern about them being a sustainable long term strategy.

Resolving the user's identity can overcome most measurement and attribution \citep{sinha2014estimating, yadagiri2015non, sinha2020attribution} issues, and considerable research has gone into stitching fragmented user behaviour \citep{saha2015probabilistic, kim2017probabilistic, jin2019node2bits, de2021session}. But these strategies rely on generic features such as IP to represent the same human user or use more detailed but private data. The first persists privacy concerns and leads to inaccurate stitching, while the second relies on first-party data from walled gardens and precludes any cross-ecosystem analysis. Industrial consortiums like "The TradeDesk
Unified ID 2.0" and "Advertising ID Consortium", are building their own unique identifiers based on people-based identifiers (Email ID,
Device). Some firms, such as Zillow, also create custom segments from their first-party data and allow advertisers access. However, concerns about scalability and possible future regulation mean this is not sustainable long-term strategy. As such \citet{kamena2021moving} propose for greater use of media mix modeling as a complementary approach to user-level attribution.

Interference related problems have been well studied in the literature \citep{hudgens_halloran08, blackwell2018make}. But
these assume strong restrictions on the structure of spillover. Recently some work has focused on accounting for general interference \citep{papadogeorgou2020causal, zigler-papadogeorgou2018-bipartite, ogburn2017causal}. However, all of these methods rely on complete knowledge of interference structure, which is impossible with our setup.

Another problem related to disappearance of cookies is that of identity fragmentation \citep{chatterjee2003modeling,miller2017economic}. \citet{lin2021identity} demonstrate treatment effect attenuation in presence of identity fragmentation. \citet{coey2016people} also provide a debiasing estimator for cookie-level estimates under similar assumptions. But both of these primarily target he loss of first-party cookies.

Our approach is instead based on conducting parallel experimentation followed by stratified aggregation. Instead of constructing user-level records, this approach works by running multiple experimentation at stratified cohorts and constructing user group-level records (aggregation). These are then shared across channels/brands to compute the desired estimate. 
Our proposed estimation method is unbiased but with a higher variance. However, by relaxing the need for constructing user-level data, it retains privacy, works without further assumptions and can achieve wider coverage. Our work borrows the idea of using multi-cluster experimentation from the work of \citet{hudgens_halloran08}. \citet{hudgens_halloran08}, which proposed a two-stage randomization procedure to account for interference. Our scenario, however is different in that the same unit receives multiple treatments instead of having exposure to treatments of other units. Furthermore, we do not make exposure-level assumptions.

\section{Problem Statement}

Let's say that our two websites have users in set $\mathbb{A}$ and $\mathbb{B}$. Note that $\mathbb{A}\cap\mathbb{B} \neq \varnothing$; so some users will visit both.
When third party cookies are available, these set of users will get a consistent view (in terms of experience/offers/ads etc); but without cookies this cannot be guaranteed, and a user may be exposed to multiple treatments.

%Further, define $|\mathbb{A} \cup \mathbb{B}| = N$, and $|\mathbb{A} \cap \mathbb{B}| = M$. 
%For simplicity we also assume that $|\mathbb{A}| = |\mathbb{B}| = \frac{N + M}{2}$ (it is easy to extend to the more general case). 
For each user $u_i \in \mathbb{A} \cup \mathbb{B}$ we have an associated feature vector $x_i$.
%We consider inference in the potential outcome setting. 
Each user $u_i \in \mathbb{A}$ can be exposed to treatments $1$ or $2$ on website $1$ . Similarly, each user $u_i \in \mathbb{B}$ can be exposed to treatments $1$ or $2$ on website $2$. If a user $u_i \in \mathbb{A} \cap \mathbb{B}$ then they have exposure to treatments from both websites. On the other hand those in $ \mathbb{A} \Delta \mathbb{B}$ are only exposed to one website and we label their other exposure on the other website as ``$0$''. Since the two brands collaborate without sharing data, our approach takes the perspective of one of the two brands (since they can only analyze visitors on their website), without loss of generality, let's consider the first brand. For users in set $\mathbb{A}$ we have 6 potential outcome variables, i.e., $Y_{1,0},Y_{2,0},Y_{1,1}, Y_{1,2},Y_{2,1}, Y_{2,2}$. Of these variables only one is observed for each user based on the treatment they received and whether they visited both websites. The problem is estimating the average treatment effect of shifting from treatment $2$ to $1$. \footnote{We make two standard assumptions in treatment effect literature, strong ignorability and positivity. A greater discussion about these is present in the supplementary material}

The challenges in estimation arises because of two main issues a) each website can only control allocation of treatments to their users; and b) the identity of the shared users is unknown. Whenever a user $u_i$ visits, the websites can only access the features $x_i$; and choose a treatment without information on whether $u_i \in \mathbb{A} \cap \mathbb{B} $ or how the other website might allocate treatment. Equivalently phrased we know for a user $u_i$ of website $1$ and allocated to treatment $1$ that we are observing one of $Y_{1,.}$, $Y_{1,1}$ or $Y_{1,2}$ but not which of these. On the other hand in standard causal inference we know for each observation unit, whether the observed outcome is $Y_1$ or $Y_2$. %Table \ref{tab:summary} summarizes the relation between the various choices and outcomes.

\begin{comment}

% \usepackage{multirow}
\begin{table}[]
\centering
\begin{tabular}{|l|l|l|l|}
\hline
                              Event   &    & \multicolumn{2}{l|}{Visited Website 1} \\ \hline
                    & Treatment   
                    & T1                & T2               \\ \hline
\multirow{2}{*}{Visited Website 2} & T3 & $Y_{13}$          & $Y_{23}$         \\ \cline{2-4} 
                                 & T4 & $Y_{14}$          & $Y_{24}$         \\ \hline
Did not visit Website 2               &    & $Y_{10}$          & $Y_{20}$         \\ \hline
\end{tabular}
\caption{Summary table for potential outcomes and treatment choices \label{tab:summary}}
\end{table}
\end{comment}

Let us make this more explicit:
For a group which has been allocated treatment $1$ by website $1$, the expected average outcome is given by:
 $$ \E[\Tilde{Y}_{1}] = (1-p)\E[ Y_{1,0}] + p ( \alpha \E[ Y_{1,1}] + (1-\alpha)\E[ Y_{1,2}]). $$
Here, $\Tilde{Y}_1$ is the random variable denoting the observed outcome of a visitor to website $1$ who is randomized to treatment $1$, $p$ is the fraction of users who are shared and hence visit both websites, while $\alpha$ is the fraction of these shared users who receive treatment $1$ on the second website as well.
Every user has the probability $1-p$ of only visiting website $1$ and hence consistently receives treatment $1$. For these users the average outcome is $\E[Y_{1,0}]$. For the rest of the users (who are $p$ fraction of the population), an $\alpha$ fraction of them are allocated to treatment $1$ by website $2$ (and hence are exposed to treatment pair $1,1$). The observed outcome on these users is $\E[Y_{1,1}]$. Similarly a $1-\alpha$ fraction of the shared users receive treatment $2$ from website and produce the average outcome $\E[Y_{1,2}]$. The observed average effect is the probability  weighted combination of all the contributions.

Furthermore by symmetry between the treatments, we can write a similar equation for the average outcome of group allocated treatment $2$ by website $1$.
%$$  \E[\bar{Y}_{2}] = (1-p)\E[ Y_{2,0}] + p ( \alpha \E[ Y_{2,1}] + (1-\alpha)\E[ Y_{2,2}]) $$ where $\bar{Y}_{2}$ is 
The observed advantage of treatment $1$ over treatment $2$, i.e., $\E[\Tilde{Y}_{1}] - \E[\Tilde{Y}_{2}]$ (which is also the standard treatment effect estimate) is given by:
$$ (1-p)\E[ Y_{1,0} - Y_{2,0} ]  + p ( \alpha \E[ Y_{1,1} - Y_{2,1} ] + (1-\alpha)\E[ Y_{1,2} - Y_{2,2} ]) $$
Next, we analyze the treatment effect if one could track the users and provide them a consistent experience.
Every user has the probability $1-p$ of only being on website 1 and hence receiving consistently either treatment 1 or 2. For these users the treatment effect is $\E[ Y_{1,0} - Y_{2,0} ]$. The rest of the users; who are $p$ fraction of the population; consistently receive treatment pair $(1,1)$ or $(2,2)$. The corresponding effect is $\E[ Y_{1,1} - Y_{2,2} ]$. The average treatment effect is the population weighted combination of the two contributions.
$$ \text{TE} = (1-p)\E[ Y_{1,0} - Y_{2,0} ]  + p ( \E[ Y_{1,1} - Y_{2,2} ]) $$
It is clear from the desired treatment effect and the observed treatment effect are mismatched due to contributions from the cross treatment outcomes $ Y_{2,1}$ and $ Y_{1,2}$ . Moreover it is clear that the mismatch increases with $p$ the fraction of shared users.

In the next section we present our method of estimating the true treatment effect.

\section{Method}

We can make more than one macroclusters with a suitable mix of treatment allocation. Each user is first assigned to a macrocluster corresponding to one of the treatment allocation strategies. Then within each such macrocluster, randomization is done for the allotment of treatment. The idea behind this approach is similar to the multi-stage randomization technique of \citet{hudgens_halloran08}.

One possible way to solve this is stratified randomization with consistency across platforms/websites. More specifically, for each user $u_i$, we put them in a category/macrocluster based on their features $x_i$, and this function or mapping the users to macroclusters is shared across all websites. One can achieve this via something similar to FLoC or Topics API, which allows advertisers a partial view of the user's preferences. Note that privacy is still maintained here; since no website shares user-specific information with the other website. 

%Another equivalent way is to consider running the experiment in multiple steps where a simple A/B experiment is conducted in each step. We shall stick to the cluster convention.

Recall that the goal of this exercise is to estimate the following estimand of interest: $(1-p)\E[ Y_{1,0} - Y_{2,0} ]  + p ( \E[ Y_{1,3} - Y_{2,4} ])$. This reflects the following notation. The treatment options at website $1$ are $T_1$ and $T_2$, and those at website $2$ are $T_3$ and $T_4$. The notation $E[Y_{1,0}]$ denotes the expectation of the outcome in visitors who have seen $T_1$ at the first website but have yet to visit website $2$. If we are in the most likely scenario where $T_1 = T_3$ and $T_2 = T_4$, our estimand boils down to  $(1-p)\E[ Y_{1,0} - Y_{2,0} ]  + p ( \E[ Y_{1,1} - Y_{2,2} ])$. To keep our solution generic, we are sticking to notation with four treatments $T_1$, $T_2$, $T_3$, and $T_4$. 

%Now consider that experiments were conducted on two such clusters. 
Using this macro-level aggregation of user information, one can do randomization with the following approach. Let's assume we have only two macroclusters, $C_1$ and $C_2$.
In the first cluster, the allocation ratio of the treatments T3 and T4, as chosen by the second website, is $\alpha \neq 0.5$; while in the second cluster, the allocation ratio is $1-\alpha$. We depict the observed average outcomes for cluster $Ci$ and treatment $j$ by the variable $\bar{Y}^{Ci}_{j}$ (an estimate of $\E(\Tilde{Y}_j^{Ci})$). Then given our setting we have the four observed outcomes $\bar{Y}^{C1}_{1}, \bar{Y}^{C1}_{2}, \bar{Y}^{C2}_{1}, \bar{Y}^{C2}_{2}$.

Our estimator of the treatment effect is given by:
\begin{align}
\begin{aligned}
\hat{\text{TE}} = \dfrac{1}{2\alpha -1 }( \alpha \bar{Y}^{C1}_{1} + (1-\alpha) \bar{Y}^{C1}_{2} - (1-\alpha) \bar{Y}^{C2}_{1} - \alpha \bar{Y}^{C2}_{2})
\end{aligned}
\end{align}

\begin{claim}
\begin{math}
{\hat{\text{TE}}}
\end{math} is an unbiased estimate of the treatment effect \text{TE} 
\end{claim}
A proof of this claim along with and estimate of the variance of this estimator can be found in Appendix \ref{apx:proof_basic}. 

\subsection{Covariate Adjustment}
Assuming the randomization is perfect, the earlier treatment effect estimate is unbiased unconditionally.
However when there is an imbalance in respect of some covariate between the groups, adjusting for the baseline effect leads to a betterestimate. This is important if the baseline covariate(s) is moderately correlated with the outcome. In such a case differences between the outcome attributable to differences in the baseline covariate can be adjusted, leading to a more efficient estimater for the treatment effect. Adjusting for covariates is often done through fitting a regression model for the outcome, with indicator variable for the randomized group and other variables as covariates. We provide a similar regression based method to improve the efficiency of our earlier estimator.

\begin{algorithm}[hbt]
  \small{
  \textbf{Input:} Vector of outcomes $Y^{C1}_{1},Y^{C1}_{2},Y^{C2}_{1},Y^{C1}_{2}$\; allocation $\alpha$ \\
  \textbf{Output:} Covariate adjusted treatment effect $\hat{\text{ATE}_{\text{cov}}}$
  \begin{enumerate}
  \item Let $Z^{Cj}_i$ = $\delta Y^{Cj}_i$ i.e. it is 1 if $Y$ corresponds to treatment 1and 0 otherwise
  \item Fit OLS($[Y^{C1}_{1}, Y^{C2}_{2}] \sim X + 1 + [Z^{C1}_{1}, Z^{C2}_{2}])$ 
  \item Let $\beta_1$ be coefficient of $Z$ as estimated in previous step
  \item Fit OLS($[Y^{C1}_{2}, Y^{C2}_{1}] \sim X + 1 + [Z^{C1}_{2}, Z^{C2}_{1}]$) 
  \item Let $\beta_2$ be coefficient of $Z$ as estimated in previous step
  \item $\hat{\text{ATE}_{\text{cov}}} = \frac{1}{2\alpha-1} ( \alpha \beta_1 + (\alpha -1) \beta_2)$
  \end{enumerate}
    \caption{Covariate Adjustment Algorithm}
    \label{alg:calibrate}
  }
\end{algorithm}

Our method is specified in Algorithm $\ref{alg:calibrate}$. We fit two linear least square estimators; and combine the regression coefficients obtained from them. We use the variable $Z$ as an indicator of treatment allocation. Unlike normal covariate adjustment where outcomes from the same group is chosen; here the model is fit from outcomes of different groups. We fit model with outcomes of treatment 1 from cluster $C1$ and outcomes of treatment 2 from cluster $C2$ and vice versa. The proof of correctness is presented in the Appendix \ref{apx:proof_reg}.
%An intuitive version of why this works is as follows:

\section{Experiments \& Demonstration}
\subsubsection{Description}
Consider the following scenario. Website 1 and its affiliate website 2, want to do a synchronous change (for example put the same banner on both websites). Ideally with third party cookies this experimented can be conducted in a standard manner. However without such identifier both websites have to individually randomize their treatments.

\subsection{Synthetic Data}
We first conduct simulation experiments where, by design, all parameters are known and adjustable. We can then quantitatively measure the performance of our method across different ranges of available parameters.

This simulation was conducted by generating 10000 observations, and each experiment was repeated 20 times. Each individual potential outcome was obtained via a noise-corrupted Gaussian distribution. Furthermore, each outcome is also influenced by a user-specific covariate. Specifically, each potential outcome variable $Y_{ij}$ at a single observation unit is obtained via a linear model from covariates $X$.
%generative process dependent on covariate $X \in \R$ and independent noise $\epsilon$  as:
%\[
%  Y_{ij} \sim \mu_{ij} + w^{(ij)} X +  \epsilon
%\]
%The variables $X$ is drawn from standard normal distribution and $\epsilon$ uniformly from $[-1, 1]$. 
The true treatment effect in this case is $(1-p)[ \mu_{1,0} - \mu_{2,0} ]  + p ( \mu_{1,3} - \mu_{2,4}) $. We vary two parameters $\delta_1 = \mu_{1,4} - \mu_{1,3}$ and $\delta_2 = \mu_{2,4} - \mu_{2,3}$ \\

\subsubsection{Results}
\begin{figure}[h]
\centering
\includegraphics[width=\linewidth]{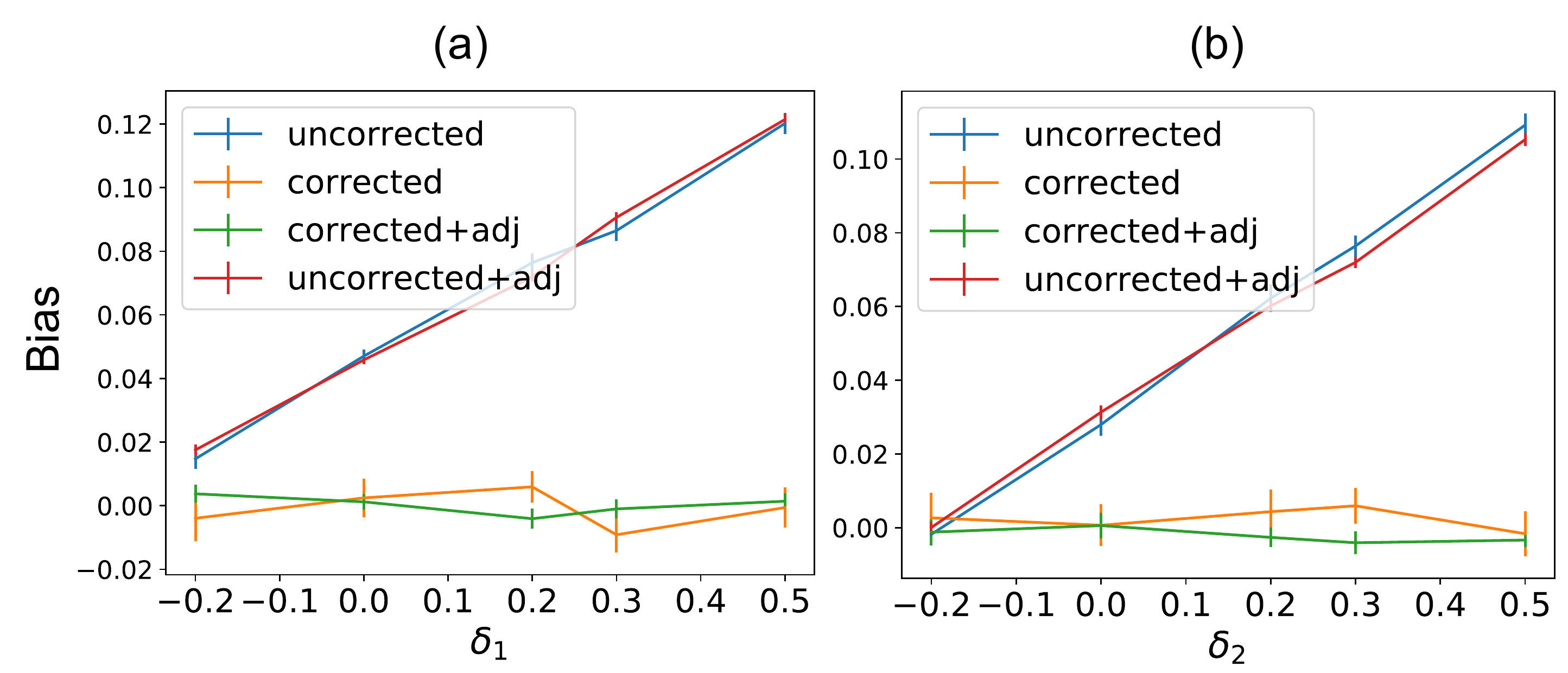}
\caption{Results on synthetic data. Std error of estimate and bias against variation of interference parameters (a) $\delta_1$ (b) $\delta_2$  \label{fig:synth}}
\end{figure}

We conducted these simulations and measured the error in the estimated $\text{ATE}$ versus the actual $\text{ATE}$ for four methods. These include the standard ATE estimate (uncorrected) , standard ATE with covariate adjustment (uncorrected + adj), our ATE estimate without (corrected)and with covariate adjustment (corrected+adj). Due to inherent variability caused by sampling, there will always remain some variability in the estimate. We present in Figure \ref{fig:synth} the results from the experiments. The plots depict both the bias of the estimator and the standard error of the estimator. The corrected method provides unbiased estimator of the true ATE while the uncorrected estimator is biased. Furthermore, as expected our proposed covariate adjustment method gives an unbiased estimate.

\subsection{Observational Data}
Next we conduct experiments from the experimental logs from Target, an experience cloud product for designing customers’ experience. Each entry in the data corresponds to a user visit, and records the users conversion event (which is our targeted binary response) along with the specific experience which the user was served and a variety of covariates such as session duration, browser details, time etc.
There are six possible experiences which were regrouped for our purposes into two treatments.

\subsubsection{Scenario 1}
We split the total time period into two halves, and considered the visitors in the two time periods as visiting two separate websites. This allowed us to simulate the visits on two websites with related treatments. A user who is present in both time periods is a user with exposure to both parts of the treatment, while a user who is present in only one time period corresponds to a user who visits only one website. Splitting the records in a such a way we can create a joint distribution of outcomes and treatments from which samples can be obtained. %As a covariate we chose the variable \emph{ENVcolor}.

\subsubsection{Scenario 2}
We also ran another experiment where we directly isolated users who got exposure directly to multiple treatments and tried estimated ATE from these experiments as well.
Since the outcomes in this case are sampled from historical logs, the effect of interference is fixed and cannot be changed. However one can analyze the effect of changing the fraction of users who receive exposure to multiple treatments.

\subsubsection{Results}
\begin{figure}[h]
\centering
\includegraphics[width=\linewidth]{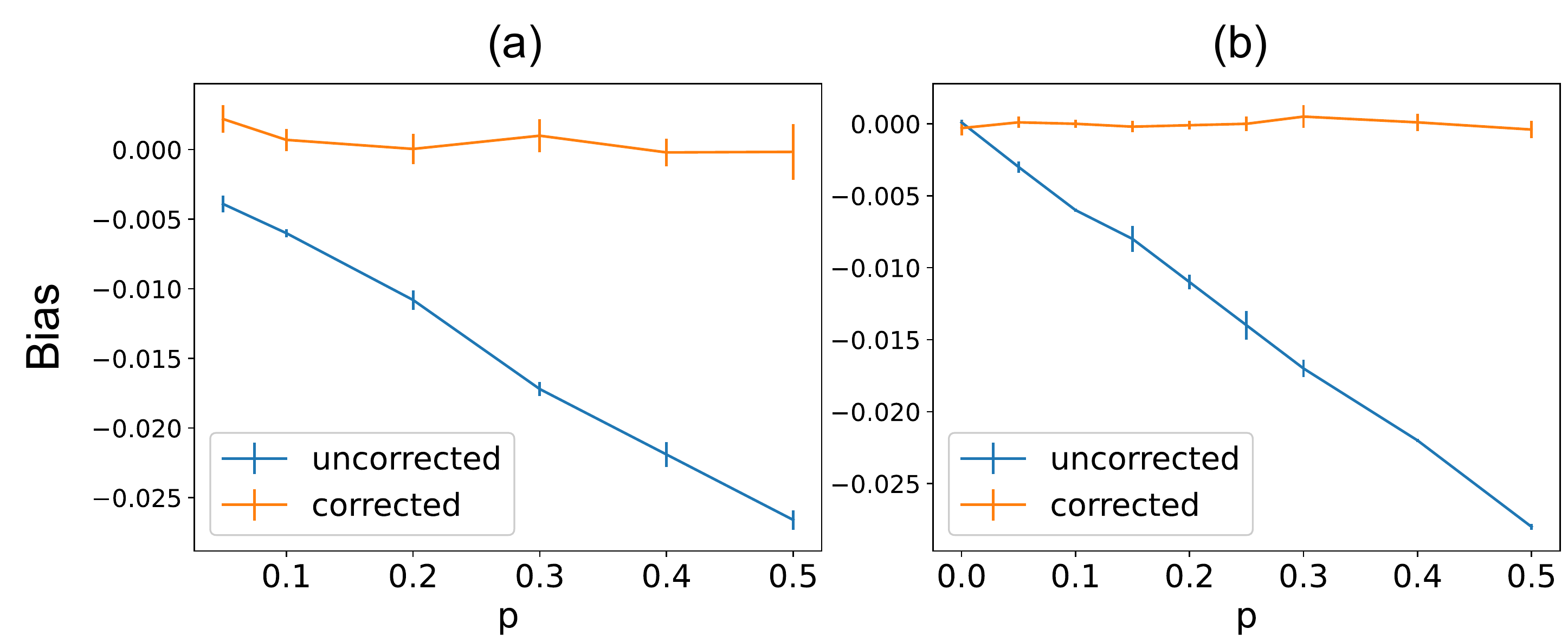}
\caption{Results on real data. Standard error of estimate and bias against the probability of shared user a) scenario 1 and b) scenario 2   \label{fig:target}}
\end{figure}

We conducted these simulations and measured the error in the estimated $\text{ATE}$ versus the true $\text{ATE}$. We present in Figure \ref{fig:target} the results from the experiments. The plots depicts both the bias of the estimator and the standard error of the estimator. The actual data has less than 0.1\% users who received exposure to different treatments, which means the bias in the historical record is minimal. As such we present the results of a simulation done by resampling the records.

%It is clear that the corrected method provides unbiased estimator of the true ATE while the uncorrected estimator is biased.  

As expected the bias of the standard estimator increases as the probability of user having multiple visits increases. This is in line with the theoretical analysis earlier, which shows that the error is proportional to $p$ ( the probability of interfered outcome).

\subsection{Demonstration Plan}
\begin{figure}[h]
\centering
\includegraphics[width=0.95\linewidth]{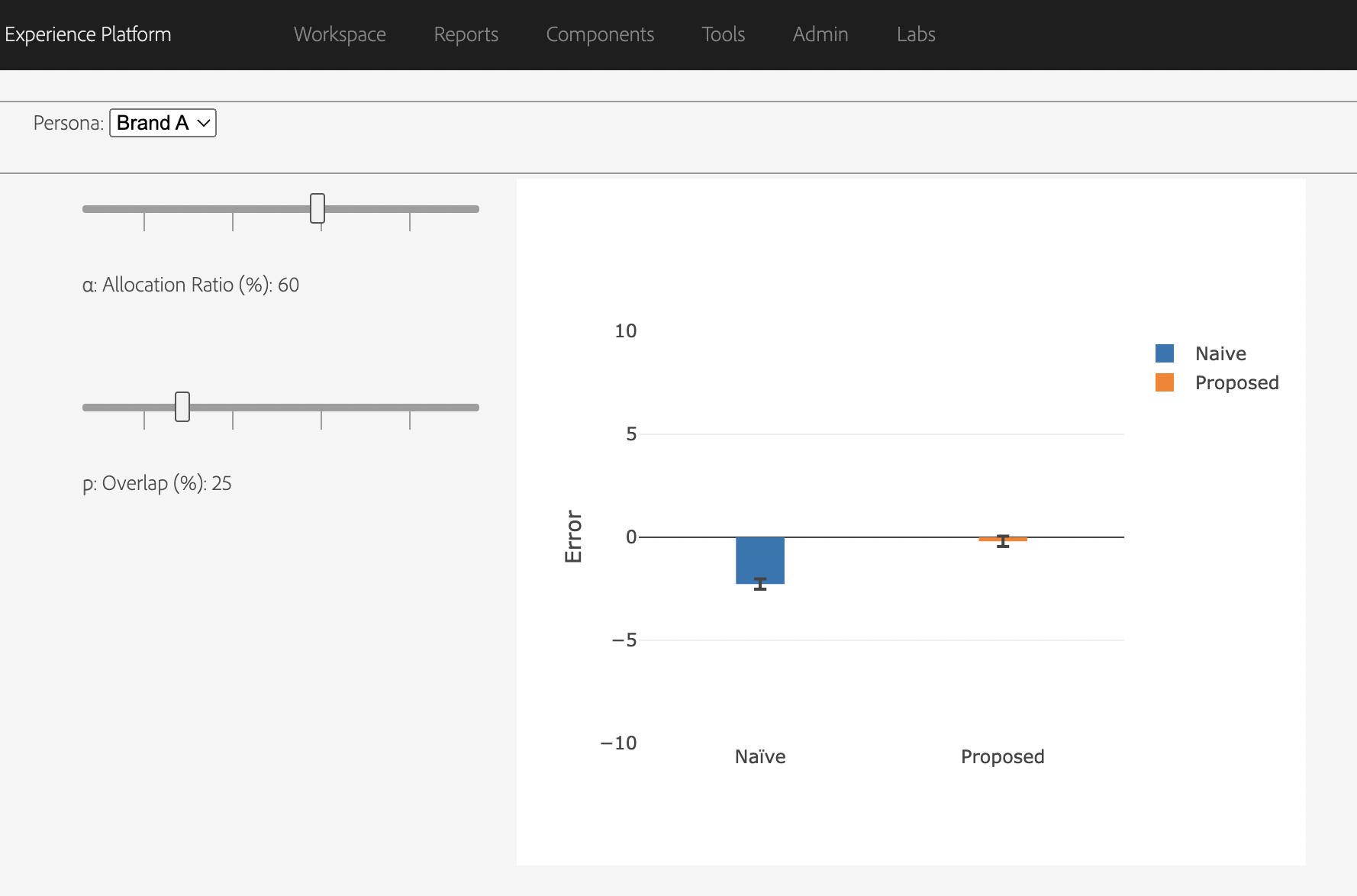}
\caption{Brand A: With $\alpha$ at 60\% and overlap of 25\% our method has less error than the naive strategy}
\label{fig:demo1}
\end{figure}

\begin{figure}[h]
\centering
\includegraphics[width=0.95\linewidth]{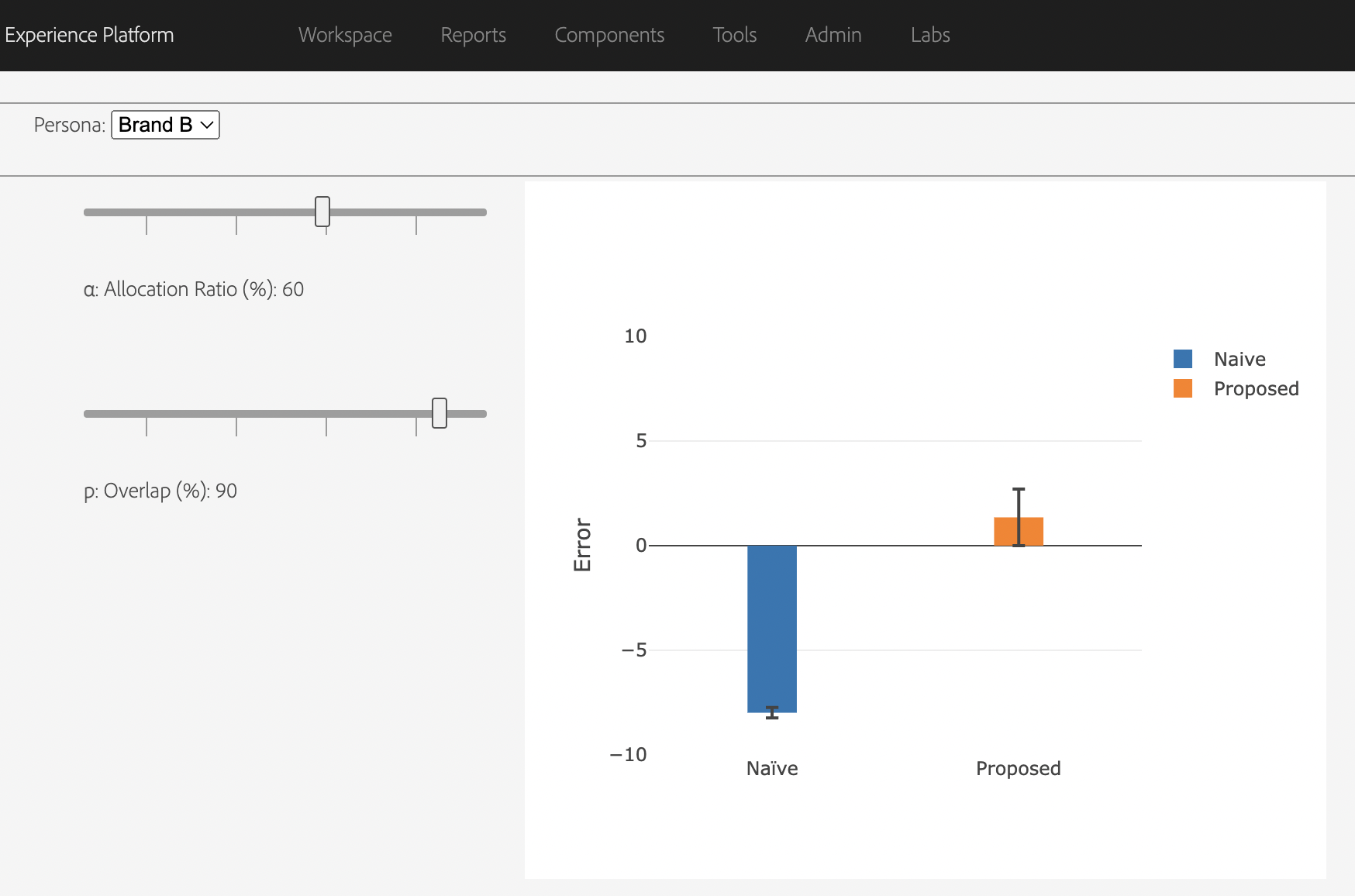}
\caption{Brand B: With $\alpha$ at 60\% and considerable overlap: 90\% our proposed method is even better than the naive strategy}
\label{fig:demo2}
\end{figure}
%\saayan{Provide a screenshot of the demo, and share some thoughts on wat will be shown}

We develop a demonstration of our technology that shows how a marketer for a brand collaborating on conducting an A/B test jointly with another brand can design and analyse an experiment by only sharing the parameters of the experiment. In Figure \ref{fig:demo1} the marketer for Brand $A$ specifies the allocation ration of $\alpha$. The chart shows the estimated treatment effect estimated from the naive difference of treatment differences, and the proposed ATE estimate. Our approach suggests that there is no treatment effect (which is the truth in the simulated scenario), while the naive estimate would suggest a negative treatment effect. Similarly, we show the experience for brand $B$ in Figure \ref{fig:demo2}.

\section{Conclusion}
We have proposed a two-stage experimental study design that two brands can use to jointly test the treatment effect of an experiment. This strategy may be used by two brands to collaborate on testing alternate experiences jointly. Our approach has no dependence on third-party cookies. More importantly, it does not require any individual level information to be shared between the two brands, thus protecting the privacy of end-customers. On the other hand, we show that the proposed ATE estimate is unbiased, thus ensuring the statistical validity of the question of interest. We also give a theoretical formula to compute the variance of our estimate. Finally, we show that in the same setting, we can compute the regression adjusted estimate of the ATE. Regression adjustment allows the A/B test to achieve higher precision in the ATE estimate, and thus being able to address this question under the challenging constraints allows us to end A/B tests quickly. Additional work is needed to address multiple treatment arms to control for multiple hypothesis testing. It will also be interesting to explore the effect of identity fragmentation on a single brand's website. Finally, our current approach treats treatments by different websites as distinct. This leads to a combinatorial explosion as the of number of treatments/websites increases. Another future research direction would be to extend this approach to handle greater degree of fragmentation. 

\bibliography{mybib}
\bibliographystyle{ACM-Reference-Format}

\appendix
\onecolumn

\section{Basic Estimator}
\label{apx:proof_basic}

Our estimator of the treatment effect is given by:
\begin{align}
\begin{aligned}
\hat{\text{TE}} = \dfrac{1}{2\alpha -1 }( \alpha \bar{Y}^{C1}_{1} + (1-\alpha) \bar{Y}^{C1}_{2} - (1-\alpha) \bar{Y}^{C2}_{1} - \alpha \bar{Y}^{C2}_{2})
\end{aligned}
\end{align}

\begin{claimapx}
\begin{math}
{\hat{\text{TE}}}
\end{math} is an unbiased estimate of the treatment effect \text{TE} 
\begin{proof}
We can see from the earlier discussion that the expected values of the average observed outcomes is given by:
\begin{align}
  \phantom{i + j + k}
  &\begin{aligned}
  \E[\bar{Y}^{C1}_{1}] = (1-p)\E[ Y_{1,0}] + p ( \alpha \E[ Y_{1,3}] + (1-\alpha)\E[ Y_{1,4}])
  \end{aligned} \nonumber \\
  &\begin{aligned}
    \E[\bar{Y}^{C1}_{2}] = (1-p)\E[ Y_{2,0}] + p ( \alpha \E[ Y_{2,3}] + (1-\alpha)\E[ Y_{2,4}])
  \end{aligned} \nonumber \\
  &\begin{aligned}
    \E[\bar{Y}^{C2}_{1}] = (1-p)\E[ Y_{1,0}] + p ( (1-\alpha) \E[ Y_{1,3}] + \alpha \E[ Y_{1,4}])
  \end{aligned} \nonumber \\
  &\begin{aligned}
    \E[\bar{Y}^{C2}_{2}] = (1-p)\E[ Y_{2,0}] + p ( (1-\alpha) \E[ Y_{2,3}] + \alpha \E[ Y_{2,4}]) 
  \end{aligned} \nonumber
\end{align}

Now by linearity of expectations:
\begin{align*}
  \E[\hat{\text{TE}}] = \dfrac{1}{2\alpha -1 }( & \alpha  \E [ \bar{Y}^{C1}_{1}] + (1-\alpha) \E[\bar{Y}^{C1}_{2}]  - (1-\alpha) \E[\bar{Y}^{C2}_{1}] - \alpha \E[\bar{Y}^{C2}_{2}])\\
\end{align*}
Plugging in the expectations from the earlier equations gives:
\begin{align*}
  \E[\hat{\text{TE}}] = (1-p)\E[ Y_{1,0} - Y_{2,0} ]  + p ( \E[ Y_{1,3} - Y_{2,4} ]) \\
\end{align*}
 which is the desired treatment effect .
\end{proof}
\end{claimapx}
The average outcome in each is a combination of the potential outcomes weighted by their relative weights. We have three outcome values, and three effects to estimate: which means that generically these can be uniquely solved.

%\begin{rem}
%The key issue here is balance:  ideally we want the category to depend only on irrelevant features; or at least a non zero probability of over any relevant feature information. \footnote{This is a bit tricky, if the individual treatment effect has a component which is independent of features; then that component is identifiable}
%\end{rem}

%\item \textbf{Kernel methods}: Optimal bipartite design in \citet{pouget2019variance} also does not consider the effect of covariates on the potential outcomes. On the other hand separation of treatment and output units means there are now further possible dependence due to presence of two sets of covariates. On the other hand majority of causal inference literature considers only mean effect estimation. This is true in the recent bipartite literature as well \citep{pouget2019variance, doudchenko2020causal}. There have been some recent forays into analysing distributional effects of treatments \citep{veitch2021counterfactual} using kernel methods. One can use kernel methods to address non-linear effects as well as distributional effects of the treatment in the bipartite setting. . This can be exceptionally useful in the journey setting, where we can have time dependent and/or non singleton effects of options

\subsection{Uncertainty Analysis}

Since the estimation process is just a linear combination of different average quantities, we can provide an easy upper bound to the variance of our estimator in terms of variance of outcomes. Let $V_M$ and $V_m$ be the maximum  and minimum variance among all outcomes i.e $V_M = \max{\text{Var}(Y_{1,0},Y_{2,0},Y_{1,3}, Y_{1,4},Y_{2,3}, Y_{2,4})}$ and $V_m = \min{\text{Var}(Y_{1,0},Y_{2,0},Y_{1,3}, Y_{1,4},Y_{2,3}, Y_{2,4})}$ .
Then, $\dfrac{V_m}{\sqrt{n}} \leq  Var(\bar{Y}^{Ci}_{j}) \leq \dfrac{V_M}{\sqrt{n}}$ where $n$ is the number of observations in each group.
and 
$$
\Var(\hat{\text{TE}}) = \dfrac{2(\alpha^2 + (1-\alpha)^2)}{(2\alpha -1)^2}\Var(\bar{Y}^{Ci}_{j}) 
$$
From above expression it is clear that the minimum variance is obtained at $\alpha =0,1$. This is not too surprising since that corresponds to an AA vs AB test across the two websites. In such a case we can with certainity find a set of users who exactly receive one treatment (i.e. either 1,3 or 2,4), and the results across the two clusters can be combined to obtain the treatment effect.

An issue however for bounds based on the above variance terms is their validity, since the estimates are only asymptotically normal.
However we also note that each average observed effect is an average of outcomes from different users. Hence under independence of each user, Chernoff-Hoeffding bound can be used to provide non-asymptotic intervals for each experiments. Since our final estimator is a linear combination of different average outcome effects, we can trivially obtain valid confidence intervals from the Hoeffding bounds of each individual experiment.

\subsection{Estimating conditional effects}
Often businesses would be interested in conditional effects as they want to focus more on users who have potentially high conversion rates. From the expression , it is obvious how to obtain conditional average treatment effect (CATE) from the data.
As long as conditional outcomes for all four groups can be obtained the earlier expression can be used to obtain the conditional effect by replacing average outcome by conditional average outcomes. These conditionals can be estimated via non-parametric regression, matching or via propensity weighted estimators.

\begin{align}
\begin{aligned}
\hat{\text{CATE}|X} = \dfrac{1}{2\alpha -1 }( & \alpha \bar{Y}^{C1}_{1}|X + (1-\alpha) \bar{Y}^{C1}_{2}|X - (1-\alpha) \bar{Y}^{C2}_{1}|X - \alpha \bar{Y}^{C2}_{2}|X)
\end{aligned}
\end{align}

\begin{claimapx}
\begin{math}
{\hat{\text{CATE}}}
\end{math} is an unbiased estimate of the conditional average treatment effect \text{CATE} 

\begin{proof}
The proof of the above statement is analogous to our earlier proof.
\end{proof}
\end{claimapx}

\section{Covariate Adjustment}
\label{apx:proof_reg}

Let the numbers of subjects randomized to experimental treatment and control be 
$n_1=\sum_{i=1}^{n}Z_i$ and
$n_1=\sum_{i=1}^{n}(1-Z_i)$, 
$n = n_0 + n_1$. 
The sample means of outcome in each group is then given by:
\begin{align*}
   \bar{Y}_1 = \dfrac{1}{n_1}\sum Z_iY_i \\
   \bar{Y}_0 = \dfrac{1}{n_0}\sum (1-Z_i)Y_i \\
   \bar{Y} = \dfrac{1}{n}\sum Y_i
\end{align*}

Then direct usage of the the OLS formula shows that
shows that the least squares estimator for $\eta_{Z}$ is

\begin{align}
\left\{
1-\frac{n^{2}}{n_{0} n_{1}}\left(n^{-1} d_{1}\right)^{T} \sum_{x x}^{-1}\left(n^{-1} d_{1}\right) \right\} ^{-1}\left\{\bar{Y}^{(1)}-\bar{Y}^{(0)}-\frac{n}{n_{0} n_{1}} \sum_{i=1}^{n}\left(Z_{i}-\bar{Z}\right) \sum_{x Y}^{T} \sum_{x X}^{-1} X_{i}\right\}
\label{eq:cov_et}
\end{align}

$$
\text { where } d_{1}=\sum_{i=1}^{n}\left(Z_{i}-\bar{Z}\right) X_{i}, \widehat{\sum}_{x y}=n^{-1} \sum_{i=1}^{n}\left(X_{i}-\bar{X}\right)\left(Y_{i}-\bar{Y}\right) \text {, and }
$$

$$
\widehat{\sum}_{x x}=n^{-1} \sum_{i=1}^{n}\left(X_{i}-\bar{X}\right)\left(X_{i}-\bar{X}\right)^{T}
$$

Now by Slutsky's theroem, we know that
$\hat{\Sigma}_{X Y} \text { and } \hat{\Sigma}_{X X}$ converge in probability to their expectation counterparts ( i.e. the cross-covariance and covariance matrices respectively). Similarly, 
$n^{2} /\left(n_{0} n_{1}\right) \stackrel{p}{\rightarrow}\{\delta(1-\delta)\}^{-1}$, and $n^{-1} d_{1} \stackrel{p}{\rightarrow} 0$.

Hence the first term in \eqref{eq:cov_et} asymptotes to 1; while the second term becomes
$$
(\bar{Y}_1 - \bar{Y}_0) - \sum (Z_i - \bar{Z}) (\sum_{XY}^T\sum_{XX}^{-1} X_i) 
$$

Now since $Z$ is independent of $X$, the expected value of the second term in the above equation is
$$
\E[\sum (Z_i - \bar{Z}) \E[(\sum_{XY}^T\sum_{XX}^{-1} X_i)] = 0* \E[(\sum_{XY}^T\sum_{XX}^{-1} X_i)] = 0
$$

Thus asymptotically the coefficient of $Z$ connverges to the standard estimator. \citet{leon2003semiparametric} details conditions under which the above estimate has a lower variance. Furthermore well known result in regression theory \citep{greene2000econometric} imply that the above estimator is consistent and asymptotically normal under entirely unrestrictive conditions.
%; normality of the outcome conditional on (Z, X), continuous outcome,or constancy of var(Y |Z, X) are not required. Indeed, the model (3) from which it is derivedneed not even be a correct representation of E(Y |Z, X) for these results to hold
%\citep{greene2000econometric,leon2003semiparametric}
%the sample proportion randomized to treatment. Then it follows from References [14, 15] that all reasonable consistent and asymptotically normal estimators for $\beta$ either can be written exactly as or are asymptotically equivalent to an expression of the form

%Y¯¯¯(1)−Y¯¯¯(0)−∑i=1n(Zi−Z¯¯¯){n−10h(0)(Xi)+n−11h(1)(Xi)},

%where h(k)(X), k = 0, 1, are arbitrary scalar functions of X.

%When h(0)(Xi) = h(1)(Xi) ≡ 0, (5) reduces to the sample mean difference Ȳ(1) −Ȳ(0), the standard “unadjusted” estimator, which is unbiased and trivially consistent for β and asymptotically normal under our general conditions.

\section{Assumptions}

\begin{ass}[Strong ignorability]\label{apx:assum:si}
Let $Z_i^j$ be a random variable denoting the treatment allocated to user $i$ on site $j$.
$$
\forall t \quad {Z}_i^j \perp Y_{i}^j(t)| \mathbf{X}^{i}
$$
\end{ass}

This assumption first explicitly introduced by \citet{rosenbaum1983assessing} is standard assumption for causal effect estimation under the potential outcomes approach. This assumption makes the covariates $X$ admissible or deconfounding. Under strong ignorability, treatment effects can be estimated without bias using propensity weighting as shown in \citet{pearl2000models}. 

%analysis of sequential treatments. This assumption means that given the observed pre-treatment variables, the treatment assignment is statistically independent of potential outcomes. It also implies that any mediators are also independent of outcomes given all observed variables.
%This means that the treatment assignment in period $t$, $\mathbf{Z}_t$, is result of perfect randomization conditional on past treatment assignments, past outcomes, and covariates that are not affected by $\mathbf{Z}_t$. Unlike the assumption of strict exogeneity in fixed effects models \citep{blackwell2013framework}, sequential ignorability disallows any unobservable confounder. Therefore, if both the outcome and the assignment process are affected by some unobservable variables (e.g., unit fixed effects), the assumption will no longer hold. 

%\citep{robins_hernan_brumback00, blackwell2013framework, blackwell2018make}.

%Note that unlike the assumption of strict exogeneity which the DID analysis or fixed effects models are built upon, no unobservable confounder is allowed to exist under sequential ignorability. Therefore, if both the outcome and the assignment process are affected by some unobservable variables (e.g., unit fixed effects), the assumption will no longer hold. 

\begin{ass}[Positivity]\label{apx:assum:bd}
$0 < P(\mathbf{Z}_i^j = t) < 1$
for all users $i$, sites $j$ and treatments $t$. 
\end{ass}
%While this is implicitly assumed in essentially all causal literature \citep{austin2011introduction,rubin2010propensity,pearl2009remarks} we make this explicit.
This requirement of positivity (also known as overlapping) ensures that every treatment allocation is possible, or equivalently each unit $i$ has a non-zero probability of being allocated any treatment. Since the different channels are going to independently allocate treatments to any user, this can be easily ensured that there is some randomization at each website.

\begin{ass}[Independent Choice]\label{apx:assum:ic}
If $A_i$ is a random variable denoting the choice of a common user to visit both websites, then $A_i  \perp Y_i^j(t) \quad \forall t$.
\end{ass}

While this is implicitly subsumed under the strong ignorability assumption, we make this explicit. The reason for this is that strong ignorability is often used instead of no unobserved confounder. In our scenario no website actually has information about $A_i$, as they do not know if an individual user will visit the other website, and hence $A$ is unobserved.

\section{Estimating treatment effect on one channel}
\begin{figure}[ht!]
\subfigure[]{
  \includegraphics[width=0.3\textwidth]{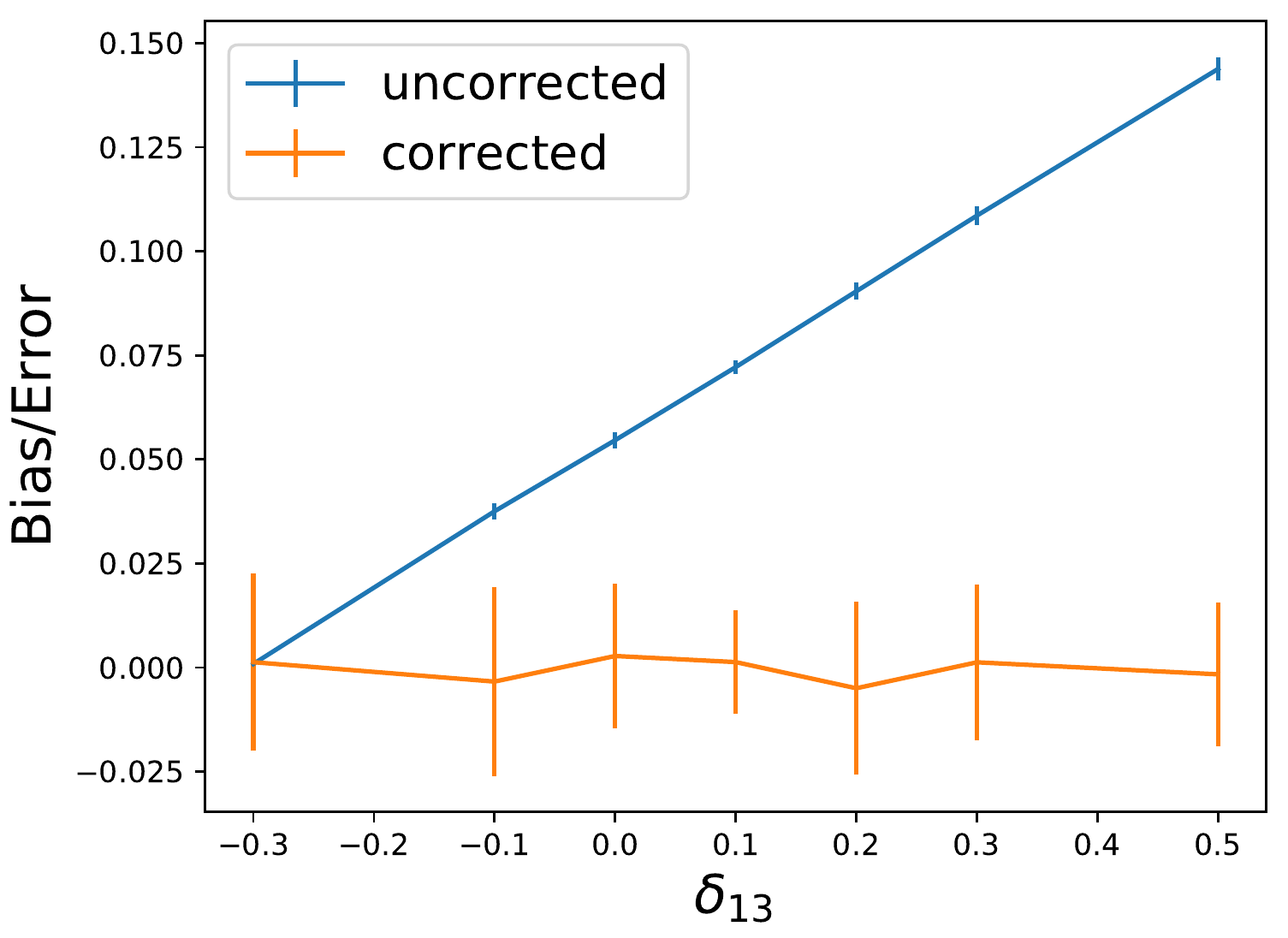}}
\subfigure[]{
  \includegraphics[width=0.3\textwidth]{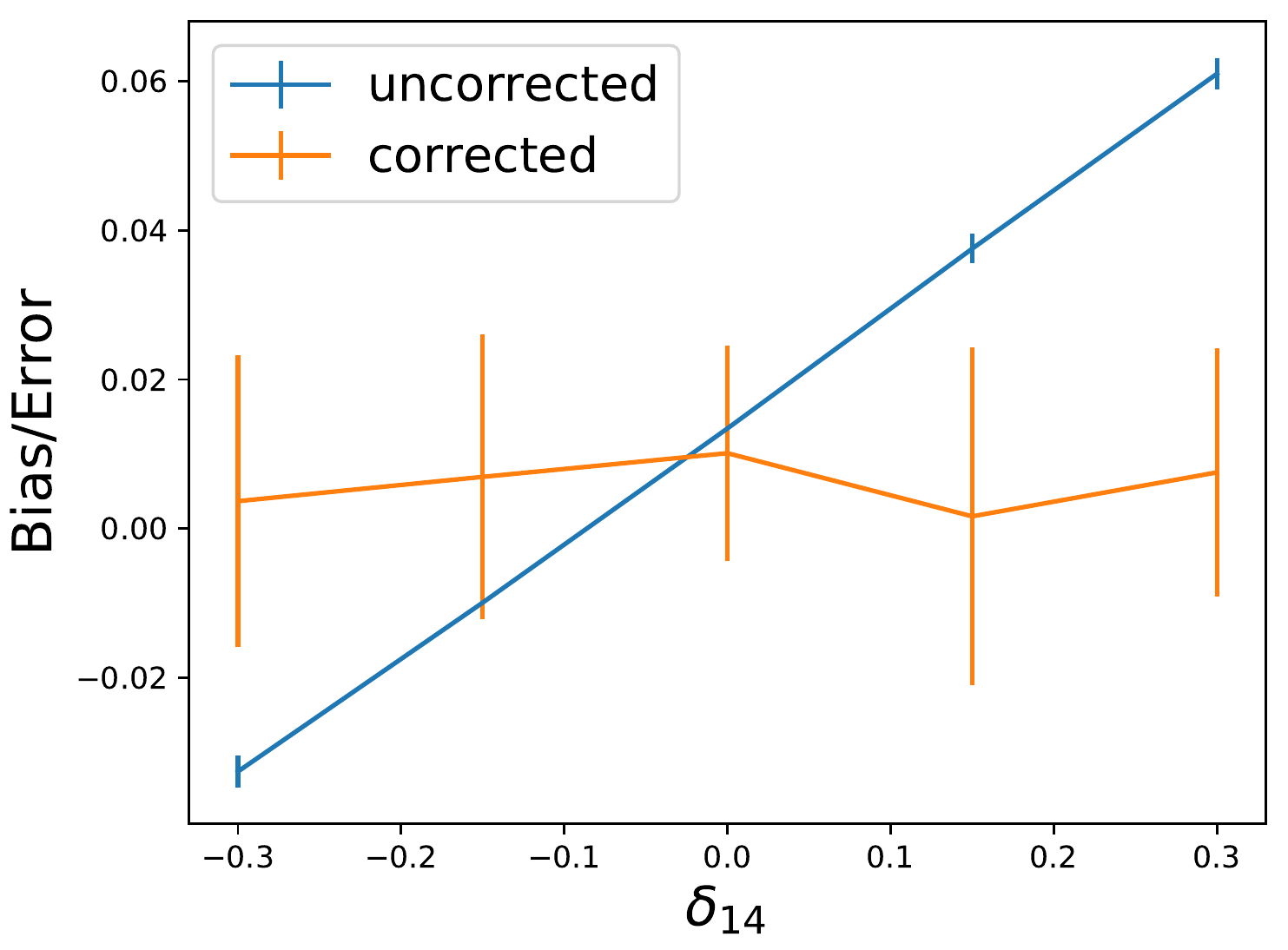}}
\subfigure[]{
  \includegraphics[width=0.3\textwidth]{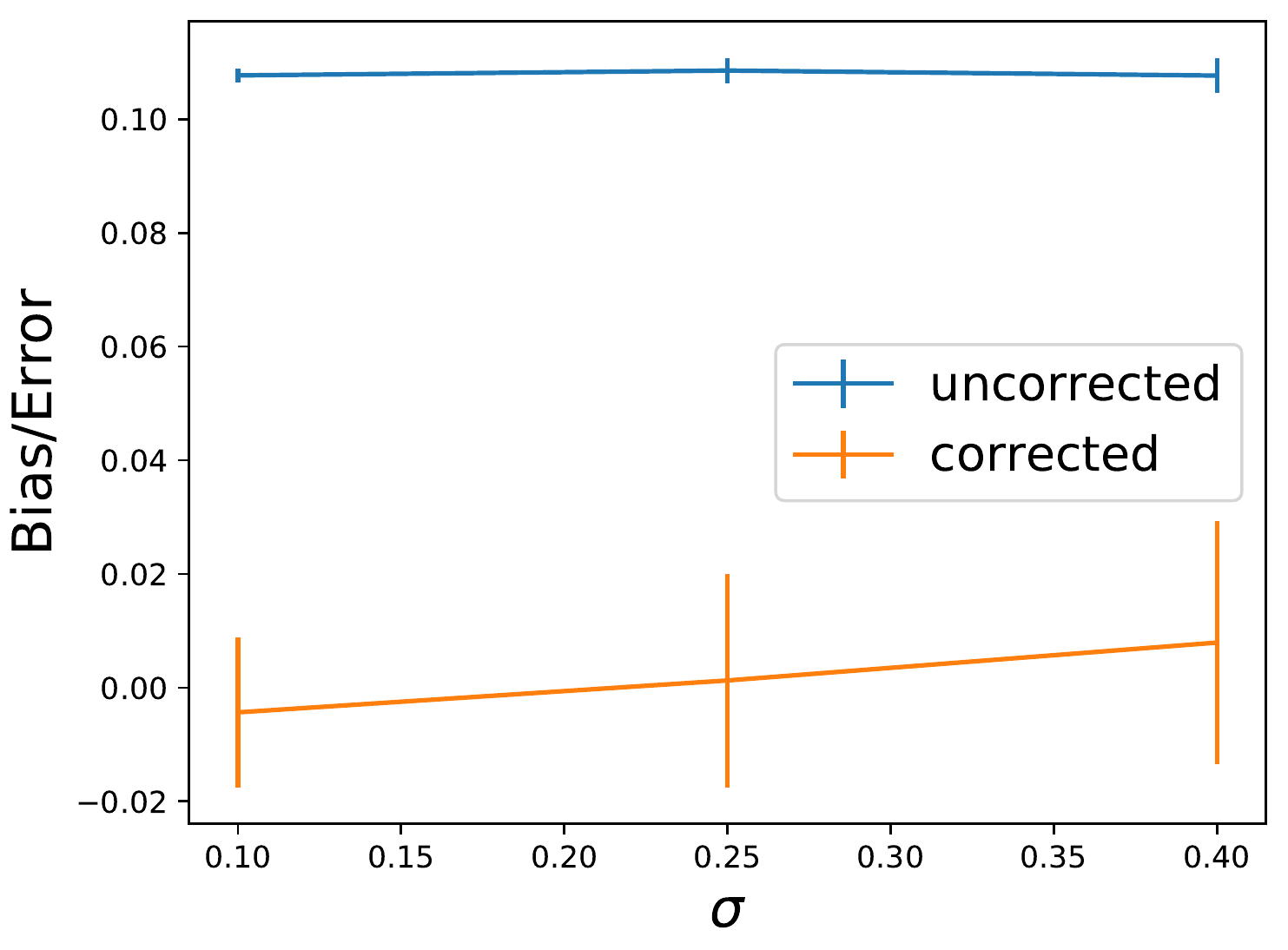}}
\caption{Results on synthetic data. Std error of estimate and bias against (a) 10 interference (b) 11 interference (c) variance of rv \label{fig:synth1}}
\end{figure}

\end{document}